\documentclass[twoside]{elsart}

\usepackage{amsmath,epsfig,amsfonts,amssymb,latexsym,subfigure,color}

\newcommand{\bc}{\begin{center}}
\newcommand{\bareret}{d}
\newcommand{\beab}{\begin{Beqnarray}}
\newcommand{\ec}{\end{center}}
\newcommand{\eeab}{\end{Beqnarray}}
 \newcommand{\SMALLCAP}	[1]	{\caption[]{ \begin{small} #1 \end{small}}}
 \newcommand{\vvec}	[1]	{\overrightarrow{#1}}
 \renewcommand{\vec}[1]{{\mathbf{#1}}}

 \newcommand{\bea}		{\begin{eqnarray}} 	
 \newcommand{\eea}		{\end{eqnarray}}
 \newcommand{\beann}		{\begin{eqnarray*}} 	
 \newcommand{\eeann}		{\end{eqnarray*}}

 \newcommand{\lrb}		{\left(}
 \newcommand{\rrb}		{\right)}

 \newcommand{\lab}		{\left\langle}
 \newcommand{\rab}		{\right\rangle}

\newcommand{\ii}{{\rm i}}
\newcommand{\isovol}{r}

\newcommand{\sep}{--}

\newcommand{\horizon}{H}

\begin{document}

 \begin{frontmatter}
 \title{Asset-asset interactions and clustering \\ in financial markets}
 \author[MPIPKS,ZiF]{Gianaurelio Cuniberti\thanksref{email}},  
 \author[MPIPKS]{Markus Porto}, and
 \author[UNIMI]{H. Eduardo Roman}
 \date{1 March 2001}
\thanks[email]		{e--mail: {\tt cunibert@mpipks-dresden.mpg.de}}
\address[MPIPKS]	{Max--Planck--Institut 
			f\"ur Physik komplexer Systeme, \\
			N\"othnitzer Stra{\ss}e 38, 
			D-01187 Dresden, Germany}
\address[ZiF]		{Zentrum f{\"u}r interdisziplin{\"a}re Forschung,
			 Universit{\"a}t Bielefeld, \\
			Wellenberg 1, D-33615 Bielefeld, Germany}
\address[UNIMI]		{Dipartimento di Fisica, Universit\`a di Milano 
			and INFN, \\
			via Celoria 16, I--20133 Milano, Italy}

\begin{abstract}
The collective phenomena of a liquid market  is characterized in terms of a particle
system scenario. This physical analogy enables us to  disentangle intrinsic features
from purely stochastic ones. The latter are the result  of environmental  changes
due to a `heat bath' acting on the  many-asset system, quantitatively described  in
terms of a time dependent effective temperature. The remaining intrinsic properties
can be  widely investigated by applying standard  methods of classical many body
systems. As an example, we consider a large set of stocks traded  at the NYSE and 
determine the corresponding asset--asset `interaction' potential. In order to
investigate in more detail the  cluster structure suggested by the short distance
behavior of the interaction potential, we perform a connectivity analysis of the
spatial distribution of the particle system. In this way, we are able to draw 
conclusions on the intrinsic cluster persistency independently of the specific
market conditions.
\end{abstract}

\begin{keyword}
Random walks, complex systems, financial markets
\\ {{\it PACS: \ }} 02.50.-r, 05.40.-a, 05.40.+j, 05.90.+m, 89.90.+n
\end{keyword}
\end{frontmatter}

The stochastic character of financial market time series is one of their distinct aspects.
Despite this random behavior, evidence has been found recently
that a certain degree of correlation is still present on extremely short  time scales
\cite{Lo91}.  The possibility to anticipate the future evolution of a single asset from the
knowledge of its past values is however minimized by the presence of traders active on
short time scales, usually with delays smaller than few seconds. 

Time dependence is just one possible domain for investigating correlation patterns inside
financial signals (see \cite{MS99} and \cite{BDLeBS96}), if contrasted with `spatial' one,
in which the features of interest are commonly referred to as {\em multivariate}
correlations. In such a spatial domain, a financial market  is seen as a complex system of
interacting constituents \cite{Fama98}, where  the study of correlations among different
assets is of peculiar importance, as in the modern theory of risk management.  On a more
fundamental level, an interesting issue is to understand how price changes can be
separated, with a sufficient degree of confidence, into {\em single asset}{\sep} and {\em
collective}{\sep} behavior.

In this paper, we are concerned with a method recently proposed to investigate
asset correlations in a stock market by means of a particle system scenario. This
can be achieved by introducing a formal map between logarithmic returns and 
distances among particles in a classical liquid \cite{CM01}. The power of this
analogy consists in the possibility of separating collective motion from the
single asset dynamics, by determining the mutual interactions among the different
assets. In this way, we can study the `thermodynamics' of the system, interpreting
its temperature as a measure of spatial volatility, which can be regarded as the
counterpart of the more familiar (temporal) volatility. The 2{\sep}asset
interaction potential is then calculated on the isothermal (isovolatile) market. 
In the following, the concept of time dependent asset{\sep}asset distance and the moving
frame of reference model of Ref.~\cite{CM01} are reviewed. The
asset{\sep}asset effective pair potential is obtained from a daily stock data taken
from the New York Stock Exchange (NYSE). The phenomenon of asset
clustering is also discussed, followed by the concluding remarks.

We consider a collection of $N$ assets from a given stock market. We wish to
define a `metric' within such a subspace of assets, so that a `distance' between
any two assets can be determined using only information of asset values at
different time horizons. The value of asset $i$ at time $t$, $\Omega_i(t)$, can be
expressed in units of the value of asset $j$, $\Omega_j(t)$, by means of the
conversion factor $P_{ij}(t)$, as $\Omega_i(t)=P_{ij}(t)\Omega_j(t)$.
By writing $\Omega_i$ as a function of $\Omega_k$, and the latter as a function of
$\Omega_j$, the  no{\sep}arbitrage equation for a liquid market is obtained
$P_{ij} = P_{ik} P_{kj}$ \cite{CM01}. The latter implicitly defines all the
cross-ratios $P_{ij}$ for any index $i,j$.

According to \cite{CM01}, the quantity  
$\bareret^\alpha_{ij}(t)=(1/{\tau_\alpha})\log[P_{ij}(t)/P_{ij}(t-\tau_\alpha)]$ can be
identified as the $\alpha$--component of a position vector between assets $i$ and $j$,
$\vec{\bareret}_{ij}$. 
It is natural to
take $\tau_{\alpha}$ from a collection of $\horizon$ time horizons, where
$\alpha\le \horizon$. 
In the $\horizon$--dimensional space, $\vec{\bareret}_{ij}$ obeys
the vector relations: ({\em a}) $\vec{\bareret}_{ii} \equiv \vec{0}$, ({\em b})
$\vec{\bareret}_{ij} = - \vec{\bareret}_{ji}$ and ({\em c}) $\vec{\bareret}_{ij} =
\vec{\bareret}_{ik} + \vec{\bareret}_{kj}$. Relation ({\em c}) results from the
non--arbitrage nature of $P_{ij}$. Using the canonical definition of the norm in an
$\horizon$--dimensional euclidean space in this case, the quantity 
$\vert\vec{\bareret}_{ij}\vert$ yields a well defined distance between assets 
$i$ and $j$.  

\begin{figure}[t]
\begin{center}
\epsfig{file=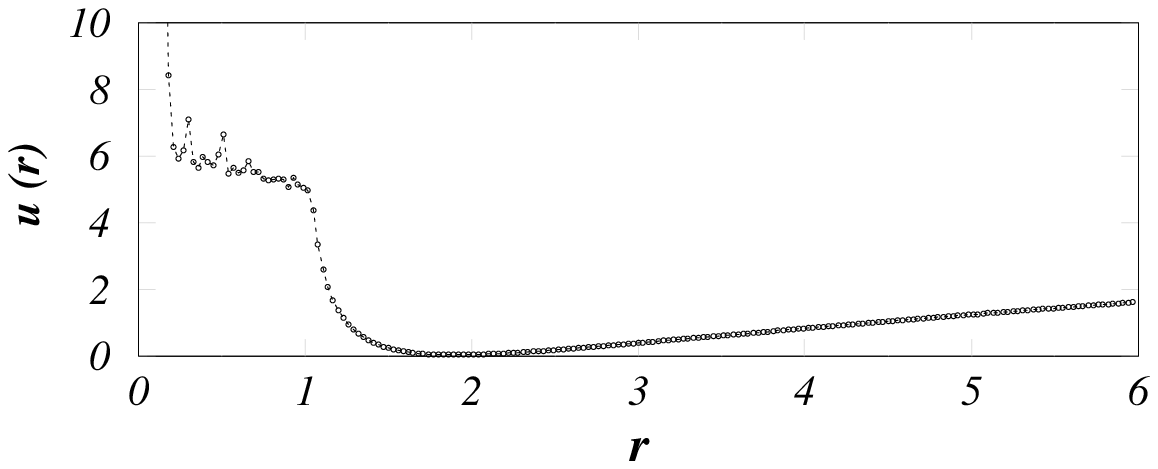, width=0.99\columnwidth} \end{center} \SMALLCAP {\label{fig:pot}
Plot of the pair potential $u(r)$ for the entire considered data set over four horizons 
($\horizon=4$) of 1, 5, 20, and 250 market days.}
\end{figure} 
In this $N$ asset system, only the relative positions between assets are meaningful.
This is due to the intrinsic character of financial markets that no asset  can be
regarded {\it a priori} as an absolute quantity. It is still possible to define an
`absolute' position of asset $i$ relative to the center of `mass' of the set, as
$\vec{x}_i \equiv (1/N) \sum_{j=1}^N \vec{\bareret}_{ij}$, with
$\vec{x}_i-\vec{x}_j=\vec{\bareret}_{ij}$, which obeys $\sum_i \vec{x}_i=0$. Note that
according to its definition, the distance between two assets is zero when the price of
one with respect to the other remains constant. 
In this way, a reference frame
can be introduced in which every single asset is assigned to an absolute position. The
problem of the $N$ assets in the market is then transformed into a physical problem of
$N$ interacting particles (a liquid) in $\horizon$ dimensions, having coordinates
$\vec{x}_1,\vec{x}_2, \dots, \vec{x}_N$. 

It turns out that the quantity $\sigma^\alpha\equiv (1/N)\sqrt{\sum_{1\le i<j \le N} 
\lrb {\bareret^\alpha_{ij}} \rrb ^2}$, is just the  standard deviation of the
$\vec{x}$'s coordinates and $(\sigma^{\alpha})^{\horizon}$  a measure of the {\em
volume} of the system. In the financial context, we refer to $\sigma^\alpha$ as the
{\em correlated volatility}, to distinguish it from the usual volatility resulting from
the {\em temporal} variability of  the assets. 
The quantities introduced so far are sufficient for solving the
eigenvalue problem for the skew symmetric matrix $\vec{\bareret}_{ij}$. One of
its three eigenvalues is zero and the corresponding eigenspace is orthogonal
to both 
$(\vec{1},\vec{1}, \dots, \vec{1})^{\rm t}$
and 
$(\vec{x}_1,\vec{x}_2, \dots, \vec{x}_N)^{\rm t}$. 
The two remaining eigenvalues are $\pm \ii N {\boldsymbol \sigma}$
corresponding to the eigenvectors $\vvec{\vec{1}} \mp \ii \vvec {\vec{\isovol}}$,
where $r^\alpha_i \equiv x^\alpha_i / \sigma^\alpha$.
Finally, we can define  (finite
difference) velocities ${\vec{v}}_i (t)= ({\vec{\isovol}_i (t) -  \vec{\isovol}_i
(t-\tau_1)} ) /{\tau_1}$, and obtain the liquid temperature  $T = \lab \lab {v}^2_i
(t)\rab_i \rab_t / H  $, and conclude that the correlated volatility is a measure of the {\em temperature} of the system~\cite{CM01}.

To elucidate the nature of the asset--asset interaction for the particle system defined so
far, we have calculated the two point correlation function, $g (r) = 2 (N (N-1))^{-1}
\sum_{i<j} \lab \delta \lrb r-  \left\vert \vec{r}_{i} (t) - \vec{r}_{j} (t)\right\vert
\rrb \rab_t$, and obtained the corresponding pair potential $u (r)$ from the relation (see
e.g. \cite{HMcD86}) $u (r) \propto - \log {g (r)}$. 

\begin{figure}[t]
\begin{center}
\epsfig{file=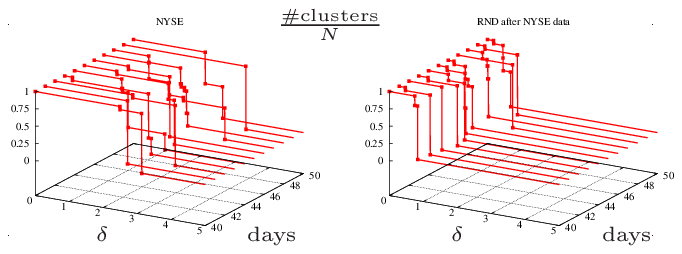, width=0.90\columnwidth}
\end{center}
\SMALLCAP{\label{fig:fig5} 
Typical behavior of the ratio of the number of distinct clusters to the total number of
particles, versus connectivity distance $\delta$ for 10 different consecutive trading days.
The left panel corresponds to real market data, and the right panel to the random
surrogates sharing the same statistical properties.}
\end{figure} 

We have determined $g(r)$ from a daily stock market data taken among 2784 equities
traded in the New York Stock Exchange (NYSE) in the period between 01-Jan-1987 and
31-Dec-1998 (3032 trading days).  To maintain a continuity of quotation, we have
selected the maximal subset of 561 assets which, in the above mentioned period,
remained  consecutively traded. 
The associated pair potential $u(r)$ is shown in Fig.~\ref{fig:pot}. 
It displays a long linear tail at large $r$
(of slope $=0.419 \pm 0.001$, and correlation coefficient$=0.9998$ calculated over 116
points in the region $2.5 < r< 6$), indicating a remarkable long range character of the
asset--asset interaction potential. These results are consistent with a previous
calculation for a by far smaller set of stocks within the german DAX index~\cite{CM01}.

We would like to learn next about the spatial heterogeneities in the $N$ particle system. 
One way to do this is to study the `clustering' of particles, in a  similar fashion as done
in the description of clusters in percolation models~\cite{percomode}.  In this case,
particles are assumed to be connected to each other if their distance is smaller than some
`connectivity distance', $\delta$.  The value of $\delta$ can be varied from $0$ to the
linear size of the whole system, $S$.  In the case of $\delta \ll S$, the particles are
clearly disconnected and one finds $N$ single clusters. By increasing $\delta$ a first
cluster is formed at a value $\delta_{\rm intra}$ giving a measure of the particle distance
within the first formed cluster. When $\delta\to S$, all particles are connected to each
other and a single cluster exists, and let $\delta_{\rm inter}$ denote the threshold to the
formation of only one cluster. For intermediate values of $\delta$, the number of clusters
decreases as $\delta$ increases.  In the case of a homogeneous particle distribution in
space, one expects to find a sort of critical value above which the number of clusters
decreases rapidly when $\delta$ is increased.   In other words, a cluster of connected
particles `percolates' the system.   However, if the system is heterogeneous the transition
may become quite different than in the homogeneous case.  First results, obtained for a
further subset of 67 NYSE stocks in the case of $H=2$ (daily and weekly horizons), suggest
that the particle system is arranged in a sort of hierarchical fashion, in which small
clusters are contained within larger ones.  This is easily seen in Fig.~\ref{fig:fig5}. The
ratio of the number of distinct clusters to the total number of particles is clearly time
dependent, alternating homogeneous and heterogeneous phases.  As a comparison, we have
generated a surrogate data set by calculating at any time step the ellipse containing real
market data in the embedding space. Its axes are the standard deviations of the fitting
two--dimensional gaussian probability out of which we have drawn the surrogate data.  It is
worth saying that, as a result, every single surrogate stock shows the stylized facts of
single asset finance (fat tails, volatility clustering~{\it etc.}) but inexorably
looses asset--asset correlation information.

\begin{figure}[t]
\begin{center}
\epsfig{file=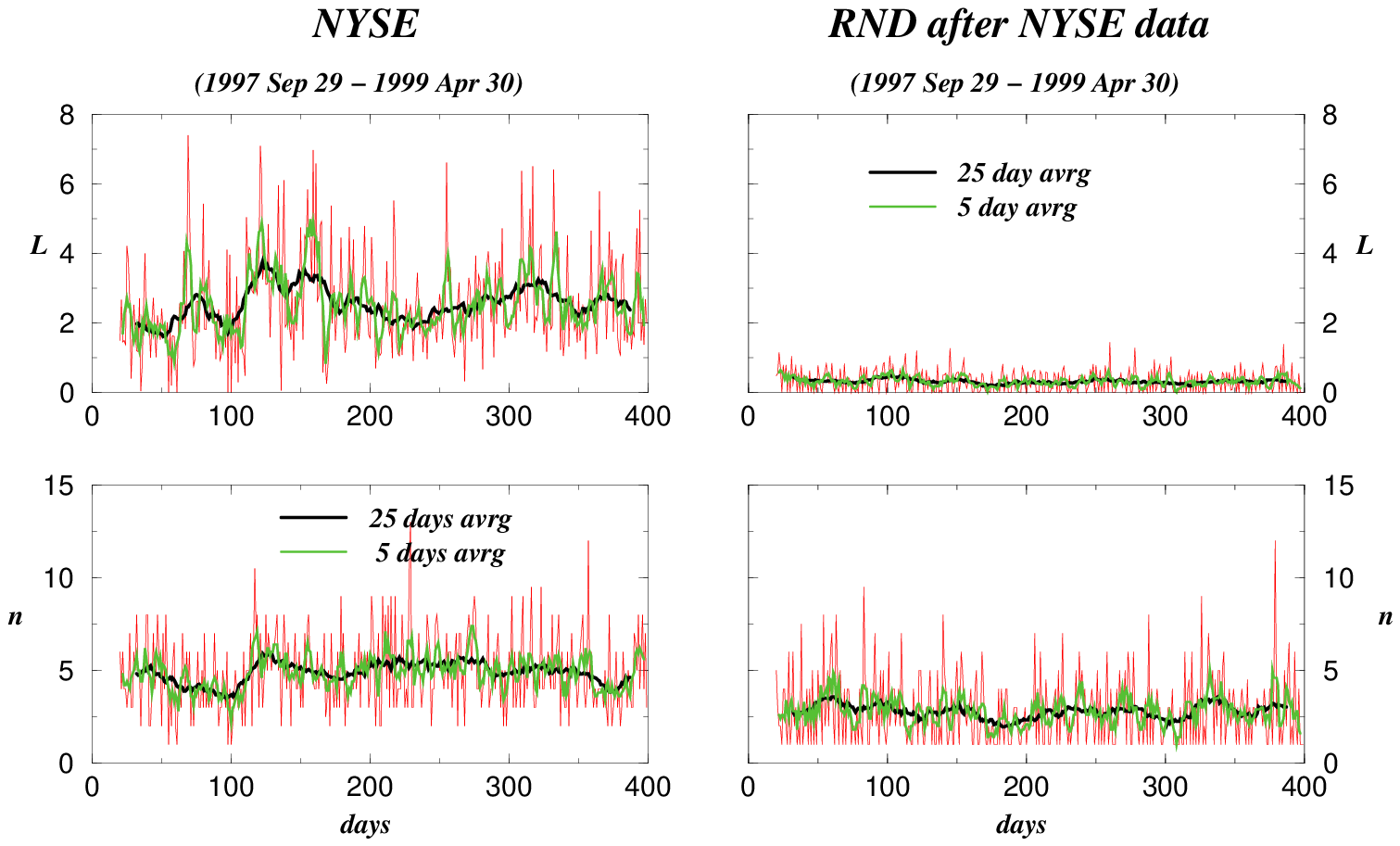, width=0.99\columnwidth}
\end{center} 
\SMALLCAP{\label{fig:fig10} 
The length of the plateux $L$ and the number of jumps $n$ of the clustering curves are
depicted for both real and random market data as in Fig.~\ref{fig:fig5}.}
\end{figure}

To capture the basic features of the staircase plot of Fig.~\ref{fig:fig5}, we have
calculated the extention of its non trivial plateaux 
$L=\delta_{\rm inter}- \delta_{\rm intra}$, and its number of jumps $n$ 
(Fig.~\ref{fig:fig10}).

Market data and random data show markedly different behaviors. Averaged market data result
in a richer structure due to the presence of  a great hierarchy of clusters at different
length scales (greater values of $n$). Moreover, at any length scale real market clusters
are also spatially better separated (greater values of $L$).  The strong time dependence of
these indicators suggests that a clustering analysis which consider a behavior on long time
scales may only give `average' answers, missing the highly non--persistent dynamics of
cluster particles.

To conclude, we have shown that a many asset market shows well marked indication of the
correlations among assets. The asset--asset potential and the hierarchical time dependent
cluster structures may reveal new paradigms for financial markets.  The same analysis has
been applied to a set of $N$ artificial prices generated with an agent based model (for
traders  with finite cash, picking stocks at random).  It comes out that, despite
individual stocks prices satisfy typical {\it single} asset empirical properties, the
calculated collective indicators presented here are in  significant disagreement with real
market results~\cite{Castiglione-private01}.

We would like to thank to F. Lillo and R. Mantegna  for providing  daily data of
equities traded in the NYSE. Fruitful discussions with Wolfgang Breymann are
also acknowledged.

\end{document}